\documentstyle[12pt]{article}
\def\cmm2{{\,\rm cm^{-2}}}
\def\cm2{{\,{\rm cm}^2}}
\def\cmm3{{\,{\rm cm}^{-3}}}
\def\gcmm3{{\,{\rm g\,cm^{-3}}}}

\def\la{\mathrel{\mathpalette\fun <}}

\def\fun#1#2{\lower3.6pt\vbox{\baselineskip0pt\lineskip.9pt
  \ialign{$\mathsurround=0pt#1\hfil##\hfil$\crcr#2\crcr\sim\crcr}}}

\begin{document}
\baselineskip=18pt
\pagestyle{empty}
\begin{center}
\bigskip


\vspace{.2in}
{\Large \bf Dark Energy and the New Cosmology}
\bigskip

\vspace{.2in}
Michael S. Turner$^{1,2}$\\

\vspace{.2in}
{\it $^1$Departments of Astronomy \& Astrophysics and of Physics\\
Enrico Fermi Institute, The University of Chicago, Chicago, IL~~60637-1433}\\

\vspace{0.1in}
{\it $^2$NASA/Fermilab Astrophysics Center\\
Fermi National Accelerator Laboratory, Batavia, IL~~60510-0500}\\

\end{center}

\vspace{.3in}
\centerline{\bf ABSTRACT}
\bigskip
A successor to the standard hot big-bang cosmology is emerging.
It greatly extends the highly successful hot big-bang model.
A key element of the New Standard Cosmology is dark energy, the causative
agent for accelerated expansion.  Dark energy is just possibly
the most important problem in all of physics.  The only laboratory
up to the task of studying dark energy is the Universe itself.

\newpage
\pagestyle{plain}
\setcounter{page}{1}
\newpage

\section{The New Cosmology}

Cosmology is enjoying the most exciting period of discovery ever.
Over the past three years a new, improved standard cosmology
has been emerging.  It incorporates the highly successful standard
hot big-bang cosmology \cite{hbb_std} and extends our
understanding of the Universe to times as early as
$10^{-32}\,$sec, when the largest structures in the Universe
were still subatomic quantum fluctuations.

This New Standard Cosmology is characterized by

\begin{itemize}

\item Flat, accelerating Universe

\item Early period of rapid expansion (inflation)

\item Density inhomogeneities produced from quantum fluctuations during inflation

\item Composition:  2/3rds dark energy; 1/3rd dark matter; 1/200th bright stars

\item Matter content:  $(29\pm 4)$\% cold dark matter; $(4\pm 1)$\% baryons;
$\sim 0.3$\% neutrinos

\end{itemize}

The New Standard Cosmology is certainly not as well established as the
standard hot big bang.  However, the evidence is mounting.

With the recent DASI observations,
the evidence for flatness is now quite firm \cite{flat}:
$\Omega_0= 1.0\pm 0.04$.
As I discuss below, the evidence for accelerated expansion
is also very strong.  The existence of acoustic peaks in the
CMB power spectrum and the evidence for a nearly scale-invariant spectrum
of primeval density perturbations ($n=1\pm 0.07$) is exactly
what inflation predicts (along with a flat Universe).  CMB anisotropy
measurements (by MAP, Planck and a host of other experiments)
as well as precision measurements of large-scale structure coming soon
from the SDSS and 2dF will test inflation much more stringently.

The striking agreement of the BBN determination of the baryon
density (from D/H measurements \cite{tytler, baryon_density})
with recent CMB anisotropy measurements \cite{flat}
make a strong case for a small baryon density compared to
the total matter density \cite{turner2001}.  The many successes
of the cold dark matter scenario  -- from
the sequence of structure formation (galaxies first, clusters
of galaxies and larger objects later) and the structure of the intergalactic
medium to its ability to reproduce the power spectrum
of inhomogeneity measured today -- makes it
clear that CDM holds much, if not all, of the truth in describing
the formation of structure in the Universe.

Cosmological measurements and observations over the next decade
or more will test (and probably refine) the New Standard Cosmology
\cite{pasp_essay}.  If we are fortunate, they
will also help us to make sense of it all.  The
most pressing item to make sense of is dark energy.
Its deep connections to fundamental physics -- a new
form of energy with repulsive gravity and possible implications for
the divergences of quantum theory and supersymmetry breaking --
put it very high on the list of outstanding problems in particle 
physics.

\section{Dark Energy}

Dark energy is my term for the causative agent of the current
epoch of accelerated expansion.  According to the second
Friedmann equation,
\begin{equation}
{\ddot R \over R} = -{4\pi G \over 3}\left( \rho + 3 p \right)
\label{eq:acc-eq}
\end{equation}
this stuff must have negative pressure, with magnitude
comparable to its energy density, in order to produce accelerated
expansion [recall $q = -(\ddot R/R)/H^2$; $R$ is the
cosmic scale factor].  Further, since
this mysterious stuff does not show its presence in galaxies
and clusters of galaxies, it must be relatively smoothly distributed.

That being said, dark energy has the following defining properties:
(1) it emits no light; (2) it has large, negative pressure,
$p_X \sim -\rho_X$; and (3) it is approximately
homogeneous (more precisely, does not
cluster significantly with matter on scales at least as large as clusters
of galaxies).  Because its pressure is comparable in magnitude
to its energy density, it is more ``energy-like'' than ``matter-like''
(matter being characterized by $p\ll \rho$).
Dark energy is qualitatively very different from dark matter.

It has been said that the sum total of progress in understanding
the acceleration of the Universe is naming the causative agent.
While not too far from the truth, there has been progress which
I summarize below.

\section{Dark Energy:  Seven Lessons}

\subsection{Two lines of evidence for an accelerating Universe}
Two lines of evidence point to an accelerating Universe.  The first
is the direct evidence based upon measurements of type Ia supernovae
carried out by two groups, the Supernova Cosmology Project \cite{perlmutter} and
the High-$z$ Supernova Team \cite{riess}.  These two teams used different
analysis techniques and different samples of high-$z$ supernovae
and came to the same conclusion:  the Universe is speeding up,
not slowing down.  

The recent discovery of a supernovae at $z=1.755$ bolsters the case
significantly \cite{riess2001} and provides the first evidence for
an early epoch of decelerated expansion \cite{turner_riess}.
SN 1997ff falls right on the accelerating Universe curve
on the magnitude -- redshift diagram,
and is a magnitude brighter than expected in a dusty open Universe
or an open Universe in which type Ia supernovae are systematically
fainter at high-$z$.

The second, independent line of evidence for the accelerating Universe
comes from measurements of the composition of the Universe, which
point to a missing energy component with negative pressure.  The argument
goes like this.  CMB anisotropy measurements indicate that
the Universe is flat, $\Omega_0=1.0\pm 0.04$ \cite{flat}.
In a flat Universe, the matter density and energy density
must sum to the critical density.  However,
matter only contributes about 1/3rd
of the critical density, $\Omega_M = 0.33\pm 0.04$
\cite{turner2001}.  (This is based upon measurements of CMB anisotropy,
of bulk flows, and of the baryonic fraction in clusters.)  Thus,
two thirds of the critical density is missing!

In order to have escaped detection this missing energy
must be smoothly distributed.
In order not to interfere with the formation of structure (by
inhibiting the growth of density perturbations)
the energy density in this component must change 
more slowly than matter (so that it was subdominant in the past).
For example, if the missing 2/3rds of critical density were smoothly
distributed matter ($p=0$), then linear density
perturbations would grow as $R^{1/2}$ rather than as $R$.  The shortfall
in growth since last scattering ($z\simeq 1100$) would be a factor of 30,
far too little growth to produce the structure seen today.

The pressure associated with the missing energy component determines
how it evolves:  
\begin{eqnarray}
\rho_X \propto R^{-3(1+w)} \nonumber\\
\rho_X /\rho_M \propto (1+z)^{3w}
\end{eqnarray}
where $w$ is the ratio of the pressure of the missing energy component
to its energy density (here assumed to be constant).
Note, the more negative $w$, the faster the ratio of missing energy
to matter goes to zero in the past.  In order to grow the
structure observed today from the density perturbations indicated by CMB anisotropy
measurements, $w$ must be more negative than about $-{1\over 2}$
\cite{turnerwhite}.

For a flat Universe the deceleration parameter today is
$$q_0 = {1\over 2} + {3\over 2}w\Omega_X \sim {1\over 2} + w$$
Therefore, knowing $w<-{1\over 2}$ implies $q_0<0$ and accelerated expansion.

\subsection{Gravity can be repulsive in Einstein's theory, but ...}

In Newton's theory mass is the source of the gravitational
field and gravity is always attractive.  In general relativity,
both energy and pressure source the gravitational field.
This fact is reflected in Eq. \ref{eq:acc-eq}.  Sufficiently large
negative pressure leads to repulsive gravity.  Thus, accelerated
expansion can be accommodated within Einstein's theory.

Of course, that does not preclude that the ultimate explanation
for accelerated expansion lies in a fundamental modification
of Einstein's theory.

Repulsive gravity is a stunning new feature of general relativity.
It leads to a prediction every bit as revolutionary as black holes  --
the accelerating Universe.  If the explanation for the accelerating
Universe fits within the Einsteinian framework, it will be an important
new triumph for general relativity.

\subsection{The biggest embarrassment in theoretical physics}

Einstein introduced the cosmological constant to balance the attractive
gravity of matter.  He quickly discarded the
cosmological constant after the discovery of the expansion
of the Universe.  Whether or not Einstein appreciated that
his theory predicted the possibility of repulsive gravity
is unclear.

The advent of quantum field theory made consideration of
the cosmological constant obligatory not optional:  The only
possible covariant form for the energy of the (quantum) vacuum,
$$ T_{\rm VAC}^{\mu\nu} = \rho_{\rm VAC}g^{\mu\nu},$$
is mathematically equivalent to the cosmological constant.
It takes the form for a perfect
fluid with energy density $\rho_{\rm VAC}$ and isotropic
pressure $p_{\rm VAC} = - \rho_{\rm VAC}$ (i.e., $w=-1$)
and is precisely spatially uniform.  Vacuum energy is almost
the perfect candidate for dark energy.

Here is the rub: the contributions of well-understood physics
(say up to the $100\,$GeV scale) to the quantum-vacuum energy
add up to $10^{55}$ times the present critical density.
(Put another way, if this were so, the Hubble time would
be $10^{-10}\,$sec, and the associated event horizon
would be 3\,cm!)  This is the well known cosmological-constant
problem \cite{weinberg,carroll}.

While string theory currently offers the best hope for
a theory of everything, it has shed precious little
light on the problem, other than to speak to the
importance of the problem. Thomas has
suggested that using the holographic principle to count
the available number of states in our Hubble volume
leads to an upper bound
on the vacuum energy that is comparable to the energy
density in matter + radiation \cite{sthomas}.  While this
reduces the magnitude of the cosmological-constant
problem very significantly, it does not solve the dark
energy problem:  a vacuum energy that is always comparable
to the matter + radiation energy density would strongly
suppress the growth of structure.

The deSitter space associated with the accelerating Universe poses
serious problems for the formulation of string theory
\cite{witten}.  Banks and Dine argue that all explanations
for dark energy suggested thus far are incompatible with
perturbative string theory \cite{dine_banks}.  At the very
least there is high tension between accelerated expansion
and string theory.

The cosmological constant problem leads to a fork in the
dark-energy road:  one path is to wait for theorists to get the
``right answer'' (i.e., $\Omega_X = 2/3$); the other path is to assume that
even quantum nothingness weighs nothing and something else
with negative pressure must be causing the Universe to speed up.
Of course, theorists follow the advice of Yogi Berra:
where you see a fork in the road, take it.

\subsection{Parameterizing dark energy:  for now, it's $w$}

Theorists have been very busy suggesting all kinds of interesting
possibilities for the dark energy:  networks of topological defects,
rolling or spinning scalar fields (quintessence and spintessence),
influence of ``the bulk'', and the breakdown
of the Friedmann equations \cite{carroll,turner2000}.
An intriguing recent paper suggests
dark matter and dark energy are connected through axion
physics \cite{barr_seckel}.

In the absence of compelling theoretical guidance,
there is a simple way to parameterize dark energy,
by its equation-of-state $w$ \cite{turnerwhite}.

The uniformity of the CMB testifies to the near isotropy
and homogeneity of the Universe.  This implies that the
stress-energy tensor for the Universe must take the perfect
fluid form \cite{hbb_std}.  Since dark energy dominates
the energy budget, its stress-energy tensor must, to a
good approximation, take the form
\begin{equation}
{T_{X}}^\mu_\nu \approx {\rm diag}[\rho_X,-p_X,-p_X,-p_X]
\end{equation}
where $p_X$ is the isotropic pressure and the desired dark
energy density is
$$\rho_X = 2.7\times 10^{-47}\,{\rm GeV}^4$$
(for $h=0.72$ and $\Omega_X = 0.66$).  This corresponds to
a tiny energy scale, $\rho_X^{1/4} = 2.3\times 10^{-3}\,$eV.

The pressure can be characterized by its ratio to the
energy density (or equation-of-state):
$$w\equiv p_X/\rho_X$$
which need not be constant; e.g., it could be a function of $\rho_X$
or an explicit function of time or redshift.  (Note, $w$ can always
be rewritten as an implicit function of redshift.)

For vacuum energy $w=-1$; for a network of topological defects
$w=-N/3$ where $N$ is the dimensionality of the defects (1 for
strings, 2 for walls, etc.).  For a minimally coupled, rolling scalar field,
\begin{equation}
w = {{1\over 2} \dot\phi^2 - V(\phi ) \over {1\over 2} \dot\phi^2 + V(\phi )}
\end{equation}
which is time dependent and can vary between $-1$ (when potential energy
dominates) and $+1$ (when kinetic energy dominates).  Here $V(\phi )$ is
the potential for the scalar field.

I believe that for the foreseeable future
getting at the dark energy will mean trying 
to measure its equation-of-state, $w(t)$.

\subsection{The Universe:  the lab for studying dark energy}

Dark energy by its very nature is diffuse and a low-energy
phenomenon.  It probably cannot be produced at accelerators;
it isn't found in galaxies or even clusters of galaxies.
The Universe itself is the natural lab -- perhaps the only
lab -- in which to study it.

The primary effect of dark energy on the Universe
is on the expansion rate.  The first Friedmann equation
can be written as
\begin{equation}
H^2(z)/H_0^2  = \Omega_M(1+z)^3 +
    \Omega_X \exp \left[ 3\int_0^z\,[1+w(x)]d\ln (1+x) \right]
\end{equation}
where $\Omega_M$ ($\Omega_X$) is the fraction of critical density
contributed by matter (dark energy) today, a flat Universe is
assumed, and the dark-energy term follows
from energy conservation, $d(\rho_X R^3) = -p_X
dR^3$.  For constant $w$ the dark energy
term is simply $\Omega_X(1+z)^{3(1+w)}$.  Note that for a flat
Universe $H(z)/H_0$ depends upon only two parameters:  $\Omega_M$ and $w(z)$.

While $H(z)$ is probably not directly measurable (however
see Ref.~\cite{Loeb}), it does affect two
observable quantities:  the (comoving) distance to an object
at redshift $z$,
$$r(z) = \int_0^z \,{dz\over H(z)},$$
and the growth of (linear) density perturbations, governed by
$$\ddot\delta_k + 2H\dot\delta_k - 4\pi G\rho_M = 0,$$
where $\delta_k$ is the Fourier component of comoving
wavenumber $k$ and overdot indicates $d/dt$.

The comoving distance $r(z)$ can be probed by
standard candles (e.g., type Ia supernovae)
through the classic cosmological observable,
luminosity distance $d_L(z) = (1+z)r(z)$.  It can also
be probed by counting objects of a known intrinsic comoving
number density, through the comoving volume
element, $dV/dzd\Omega = r^2(z)/H(z)$.

Both galaxies and clusters of galaxies have been suggested
as objects to count \cite{count}.  For each,
their comoving number density evolves (in the case of
clusters very significantly).  However, it is believed that
much, if not all, of the evolution can be modelled through
numerical simulations and semi-analytical calculations
in the CDM picture.  In the case of clusters, evolution is
so significant that the number count test probe is affected
by dark energy through both $r(z)$ and the growth of perturbations,
with the latter being the dominant effect.

The various cosmological approaches to ferreting
out the nature of the dark energy have been studied extensively
(see other articles in this {\em Yellow Book}).
Based largely upon my work with Dragan Huterer \cite{ht},
I summarize what we know about the efficacy of the cosmological
probes of dark energy:

\begin{itemize}

\item Present cosmological observations prefer $w=-1$,
with a 95\% confidence limit $w < -0.6$ \cite{perlmutteretal}.

\item Because dark energy was less important in the past,
$\rho_X/\rho_M \propto (1+z)^{3w}\rightarrow 0$ as $z
\rightarrow \infty$, and the Hubble
flow at low redshift is insensitive to the composition of
the Universe, the most sensitive
redshift interval for probing dark energy is $z=0.2 - 2$
\cite{ht}.

\item The CMB has limited power to probe $w$ (e.g.,
the projected precision for Planck is $\sigma_w = 0.25$)
and no power to probe its time variation \cite{ht}.

\item A high-quality sample of 2000 SNe distributed from
$z=0.2$ to $z=1.7$ could measure $w$ to a precision $\sigma_w
=0.05$ (assuming an irreducible error of 0.14 mag).
If $\Omega_M$ is known independently to better
than $\sigma_{\Omega_M} = 0.03$, $\sigma_w$ improves by
a factor of three and the rate of change of $w^\prime =
dw/dz$ can be measured to precision $\sigma_{w^\prime} = 0.16$
\cite{ht}.

\item Counts of galaxies and of clusters of galaxies may have
the same potential to probe $w$ as SNe Ia.  The
critical issue is systematics (including the evolution of
the intrinsic comoving number density, and the ability to identify
galaxies or clusters of a fixed mass) \cite{count}.

\item Measuring weak gravitational lensing by large-scale
structure over a field of 1000 square degrees (or more)
could have comparable sensitivity to $w$ as type Ia supernovae.
However, weak gravitational lensing does not appear to be a good
method to probe the time variation of $w$ \cite{dragan}.  The systematics
associated with weak gravitational lensing have not yet been studied
carefully and could limit its potential.

\item Some methods do not look promising in their ability
to probe $w$ because of irreducible systematics (e.g.,
Alcock -- Paczynski test and strong gravitational lensing
of QSOs).  However, both could provide important independent
confirmation of accelerated expansion.

\end{itemize}

\subsection{Why now?:  the Nancy Kerrigan problem}

A critical constraint on dark energy is that it not interfere
with the formation of structure in the Universe.  This means
that dark energy must have been relatively unimportant in the
past (at least back to the time of last scattering, $z\sim 1100$).
{\em If} dark energy is characterized by constant $w$, not
interfering with structure formation can be quantified as:
$w\la -{1\over 2}$ \cite{turnerwhite}.  This means
that the dark-energy density evolves more slowly than
$R^{-3/2}$ (compared to $R^{-3}$ for matter) and implies
\begin{eqnarray*}
\rho_X/\rho_M & \rightarrow & 0\qquad{\rm for\ }t\rightarrow 0 \\
\rho_X/\rho_M & \rightarrow & \infty \qquad{\rm for\ }t\rightarrow \infty \\
\end{eqnarray*}

That is, in the past dark energy was unimportant and in the
future it will be dominant!  We just happen to live at the
time when dark matter and dark energy have comparable densities.
In the words of Olympic skater Nancy Kerrigan, ``Why me?  Why now?''

Perhaps this fact is an important clue to unraveling the nature
of the dark energy.  Perhaps not.  And God forbid, it could be
the basis of an anthropic explanation for the size of the
cosmological constant.

\subsection{Dark energy and destiny}

Almost everyone is aware of the connection between the shape
of the Universe and its destiny:  positively curved recollapses,
flat; negatively curved expand forever.  The link between
geometry and destiny depends upon a critical assumption: that
matter dominates the energy budget (more precisely, that all components
of matter/energy have equation of state $w> -{1\over 3}$).
Dark energy does not satisfy this condition.

In a Universe with dark energy the connection between geometry
and destiny is severed \cite{kraussturner}.  A flat Universe
(like ours) can continue expanding exponentially forever
with the number of visible galaxies diminishing to a few
hundred (e.g., if the dark energy is a true cosmological constant);
the expansion can slow to that of a matter-dominated
model (e.g., if the dark energy dissipates and becomes sub
dominant); or, it is even possible for the Universe to recollapse
(e.g., if the dark energy decays revealing a negative cosmological constant).
Because string theory prefers anti-deSitter space, the third
possibility should not be forgotten.

Dark energy holds the key to understanding our destiny!

\section{The Challenge}

As a New Standard Cosmology emerges, a new set questions arises.
(Assuming the Universe inflated) What is physics underlying inflation?  What is
the dark-matter particle?  How was the baryon asymmetry produced?
Why is the recipe for our Universe so complicated?
What is the nature of the Dark Energy?  All of these questions have
two things in common:  making sense of the New
Standard Cosmology and the deep connections they reveal between
fundamental physics and cosmology.

Of these new, profound cosmic questions, none is more important
or further from resolution than the nature of the dark energy.
Dark energy could well be the number one problem in all of
physics and astronomy.

The big challenge for the New Cosmology is making sense of dark energy.

Because of its diffuse character, the Universe is likely the lab
where dark energy can best be attacked (though one should not
rule other approaches -- e.g., if the dark energy involves a light
scalar field, then there should be a new long-range force \cite{carroll2}).

While type Ia supernovae look particularly promising -- they have
a track record and can in principle be used to map out
$r(z)$ -- there are important open issues.
Are they really standardizable candles?  Have they evolved?  Is the
high-redshift population the same as the low-redshift population?

The dark-energy problem is important enough that pursuing complimentary
approaches is both justified and prudent.  Weak-gravitational lensing
shows considerable promise.  While beset by important issues involving number
evolution and the determination of galaxy and cluster masses \cite{count},
counting galaxies and clusters of galaxies should also be pursued.

Two realistic goals for the next decade are the determination of
$w$ to 5\% and looking for time variation.  Achieving either
has the potential to rule out a cosmological constant:
For example, by measuring a significant time
variation of $w$ or by pinning $w$ at $5\sigma$ away from $-1$.
Such a development would be a remarkable, far reaching result.

After determining the equation-of-state of the dark energy, the
next step is measuring its clustering properties.
A cosmological constant is spatially constant; a rolling
scalar field clusters slightly on very large scales \cite{friemanetal}.
Measuring its clustering properties will not be easy,
but it provides an important, new window on dark energy.

We do live at a special time:  There is still enough light in the Universe
to illuminate its dark side.

\paragraph{Acknowledgments.}  I thank Eric Linder for useful
comments.  This work was supported by
the DoE (at Chicago and Fermilab) and by the NASA (at Fermilab
by grant NAG 5-7092).

\end{document}